\documentclass[aps,pra,floatfix]{revtex4}
\usepackage{epsfig}\usepackage{amsmath}
\usepackage{amsmath,amsfonts,amssymb,graphics,graphicx,epsfig,color,times,bbm}

\begin{document}

\preprint{APS/123-QED}

\title{Coherent states superpositions in cavity quantum electrodynamics with trapped ions}

\author{F. L. Semi\~ao}
\author{A. Vidiella-Barranco}%
\affiliation{%
Instituto de F\'\i sica ``Gleb Wataghin'' - Universidade Estadual de Campinas, 13083-970 Campinas S\~ao Paulo Brazil}
\date{\today}

\begin{abstract}
We investigate how superpositions of motional coherent states
naturally arise in the dynamics of a two-level trapped ion coupled
to the quantized field inside a cavity. We extend our considerations
including a more realistic set up where the cavity is not ideal and
photons may leak through its mirrors. We found that a detection of a
photon outside the cavity would leave the ion in a pure state. The
statistics of the ionic state still keeps some interference effects that might be
observed in the weak coupling regime.
\end{abstract}

\maketitle
\section{Introduction} There has been a great deal of
interest in the coherent manipulation of simple quantum systems
\cite{blatt_rev,demille,haroche_rev}, mainly to the high degree of
control necessary for the implementation of quantum information
processing tasks \cite{cirac_zoller,zoller2,molmer}. In particular,
the study of trapped ions interacting with laser beams has attracted
much attention due to the significant experimental advances in the
generation of quantum states in such a system \cite{wingen,blagen}.
The interaction of trapped ions with laser beams is well understood
in terms of a semiclassical model with the electromagnetic field
being treated as a c-number, but new features mikght be revealed due
to the field quantization. The entanglement between photons and ions
is a remarkable consequence of that quantization and its potential
applications heve been motivating the experimental work in cavity
quantum electrodynamics with trapped particles \cite{blacav}. For
instance, there have been reported schemes for the generation of
specific entangled states such as Greenberger-Horne-Zeilinger (GHZ)
states \cite{knight} as well as Bell states \cite{bell}.

One of the reasons for interest in studying and experimentally coupling photons
with material particles comes from the fact that in order for the
quantum information processing to be used in its full extent, one
should be able to inter-convert stationary and flying qubits and
also faithfully transmit the flying qubits between given
positions. Those two statements are part of what it are know as
DiVincenzo's requirements for the physical implementation of quantum
computation and information \cite{Divincenzo}. The entanglement
present in the system consisting of cavities and trapped ions may be
useful in the propagation of information carried by photons between
two distant locations \cite{networks}.

It is not just entangled states involving either two level systems
or Fock states of the electromagnetic field that find applications
in quantum information. Another interesting class of nonclassical
states with high potential applications is the one formed by linear
superpositions of coherent states. This class of states has been considered
for quantum teleportation \cite{hirota_tele,xiao_tele}, logic gates
implementation \cite{kim1,kim2} and tests of local realism
\cite{sanders_local}, for instance. In this paper, we show that
superpositions of  motional coherent states may be generated in the
framework of cavity electrodynamics with trapped ions by letting the
system evolve in the resonant carrier dynamics and by performing a
measurement of the internal state of the ion. We apply the formalism
of quantum jumps to study the non ideal case including damping in
the cavity and show that the detection of photons outside the cavity
could be used to generate nonclassical states of the motion of the
trapped ion. More precisely, we show that the statistics of the
generated state keeps track of the coherence displayed on the
oscillatory behavior of the phonon number distribution and the
variation of its width from Poissonian to sub or super-Poissonian \cite{phase}.
Although that is not a deterministic protocol (it depends on
the random event of the leaking of a photon from the cavity), it might be
of interest because it could be implemented in current experimental
systems. Experiments involving trapped ions and optical
cavity fields have been performed only in the weak coupling regime
in which the cavity damping is stronger than the ion-cavity coupling
\cite{blacav}.
\section{Model Hamiltonian}
In this work we consider a single two-level ion trapped in a Paul
trap and placed inside an optical cavity. The cavity mode couples to
the ionic internal degrees of freedom \{$|e\rangle$,$|g\rangle$\}
and the system Hamiltonian is given by \cite{zeng}
\begin{eqnarray}
\hat{H}&=&\hbar\nu \hat{a}^{\dagger}\hat{a} +
\hbar\omega\hat{b}^{\dagger}\hat{b}
+\hbar\frac{\omega_0}{2}\hat{\sigma}_z \nonumber\\&&+ \hbar
g(\hat{\sigma}_+ + \hat{\sigma}_-)(\hat{b}^{\dagger}+
\hat{b})\cos\eta(\hat{a}^{\dagger}+\hat{a}), \label{H}
\end{eqnarray}
where $\hat{a}^{\dagger}(\hat{a})$ denotes the creation (annihilation) operator of the center-of-mass vibrational
motion of the ion (frequency $\nu$), $\hat{b}^{\dagger}(\hat{b})$ is the creation (annihilation) operator of photons
in the field mode (frequency $\omega$), $\hat{\sigma}$ operators are the usual Pauli matrices for the two internal
levels of the ion, $\omega_0$ is the atomic frequency transition, $g$ is the ion-field coupling constant,
and $\eta=2\pi a_0/\lambda$ is the Lamb-Dicke parameter, being $a_0$ the amplitude of the harmonic motion and
$\lambda$ the wavelength of the cavity field.

For our purposes here we may work in the Lamb-Dicke regime
($\eta\ll 1$), i.e., the situation in which the spatial
extend of the motion of the trapped ion is much smaller than the
wavelengh of the cavity field. In this regime, we may perform an
approximation that simplifies the original
Hamiltonian (\ref{H}) as follows
\begin{equation}
\cos\eta(\hat{a}^{\dagger}+\hat{a})\approx 1-\frac{\eta^2(1+
2\hat{a}^{\dagger}\hat{a})}{2}-\frac{\eta^2(\hat{a}^\dagger{}^2+\hat{a}^2)}{2}.
\label{expan}
\end{equation}
If we tune the light field so that it exactly matches the atomic
transition, i.e., $\omega_0-\omega=0$ (carrier transition), we
obtain the interaction Hamiltonian in the Lamb-Dicke
regime, which, after discarding rapidly oscillating terms reads
\begin{equation}
\hat{H}_I= \hbar g \left[1-\frac{\eta^2(1+2\hat{a}^{\dagger}\hat{a})}{2}\right]
(\hat{\sigma}_- \hat{b}^{\dagger} + \hat{\sigma}_+ \hat{b}).
\label{hamilint}
\end{equation}
The resulting Hamiltonian in equation (\ref{hamilint}) is similar to
the Jaynes-Cummings Hamiltonian but having an effective coupling
constant which in our case depends on the excitation number of the
ionic oscillator, $\hat{m}=\hat{a}^{\dagger}\hat{a}$. Such a
dependence on the intensity has already been demonstrated
\cite{inten} to be related to the occurrence of super-revivals
(revivals taking place at long times) of the atomic inversion.

\section{Results}
\subsection{Generation of superpositions of motional states}
We now consider that the system is initially prepared in a way that
the ion is in its excited state $|e\rangle$ (internal level), the
cavity in the vacuum state $|0\rangle_c$, and the vibrational motion
in the coherent state $|\alpha\rangle_v$, i.e.
$|\psi(0)\rangle=|\alpha\rangle_v|0\rangle_c|e\rangle$. Under the
Hamiltonian (\ref{hamilint}), the state $|\psi(0)\rangle$ evolves to
\begin{eqnarray}
|\psi(t)\rangle&=&\cos\left(gt\left[1-\eta^2(1+2\hat{a}^{\dagger}\hat{a})/2 \right]\right)|\alpha\rangle_v|0\rangle_c|e\rangle\nonumber\\
&&-i\sin\left(gt\left[1-\eta^2(1+2\hat{a}^{\dagger}\hat{a})/2 \right]\right)|\alpha\rangle_v|1\rangle_c|g\rangle.\nonumber\\
\label{ppsi}
\end{eqnarray}
We still have to apply the functions of the operator $\hat{a}^{\dagger}\hat{a}$  in the coherent state
$|\alpha\rangle_v$. This may be easily done by moving to the Fock basis and the result is given by
\begin{eqnarray}
|\psi(t)\rangle&=&[\cos(\omega_{\eta}t)\,|\Phi_+\rangle_v-i\sin(\omega_{\eta}t)\,|\Phi_-\rangle_v]\,
|0\rangle_c|e\rangle \nonumber\\
&&+\,[\cos(\omega_{\eta}t)\,|\Phi_-\rangle_v-i\sin(\omega_{\eta}t)\,|\Phi_+\rangle_v]\,|1\rangle_c|g\rangle,\nonumber\\
\label{ent1}
\end{eqnarray}
where $\omega_{\eta}\equiv g(1-\eta^2/2)$ and we $|\Phi_\pm\rangle_v$ are general superpositions of coherent
states given by
\begin{equation}
|\Phi_\pm\rangle_v\equiv\frac{|\alpha\:e^{i\phi}\rangle_v\pm|\alpha\:e^{-i\phi}\rangle_v}{2},
\label{ent2}
\end{equation}
where we defined the time dependent real phase $\phi=\eta^2gt$. The
state (\ref{ent1}) is an entangled state involving superpositions of
motional coherent states of the trapped ion, its internal electronic
states, and Fock states of the cavity field. It is noteworthy that
for interaction times given by $t_k=k\pi$, with $k$ being an integer
number, the state of the system reduces to
\begin{equation}
|\psi(t)\rangle=|\Phi_+\rangle_v|0\rangle_c|e\rangle +|\Phi_-\rangle_v|1\rangle_c|g\rangle.
\label{ent3}
\end{equation}
One could then obtain a disentangled motional state by performing a
measurement on the internal state of the ion. The experimental
discrimination between the two electronic levels may be done using
the very efficient electron shelving method \cite{shelving}.
Depending on the measurement outcome, the collapsed motional state
may be either $|\Phi_+\rangle_v$ or $|\Phi_-\rangle_v$. One of the
main interesting characteristics of those superposition states is
that their statistics are strongly sensitive to the value of the
phase $\phi$. The trivial case takes place when $\phi=0$, what leads
the distribution $P_m=|\langle m|\Phi_\pm\rangle_v|^2$ (phonon
statistics) to be Poissonian. However, it is well known that there
are domains in which it can be either sub or super-Poissonian. As
pointed out in \cite{phase}, when the statistics is
super-Poissonian, the distribution $P_m$ displays an oscillatory
behavior, being this a direct consequence of interference in phase
space. Such a behavior is analogous to the oscillatory photon
statistics of highly squeezed states \cite{osc}. Although similar
superposition states may also be generated using classical fields
\cite{gerry}, the possibility of entanglement with light is a unique
feature related to the quantum nature of the electromagnetic field.

The scheme proposed here relies on a not very demanding initial
preparation of the system. It requires the initial field to be in
the vacuum state $|0\rangle_c$, i.e., there is no need to prepare or
inject a coherent field state into the cavity. Additionally, the
vibrational motion of the ion has to be prepared in a coherent state
$|\alpha\rangle_v$, whose experimental realization for a
$^{9}\mathrm{Be}^{+}$ ion trapped in a RF (Paul) trap has been
already reported \cite{wingen}. Regarding the internal ionic states,
they need to be prepared in the excited state which can be achieved
by the application of laser pulses, for instance.

\begin{figure}
\includegraphics[width=8.cm]{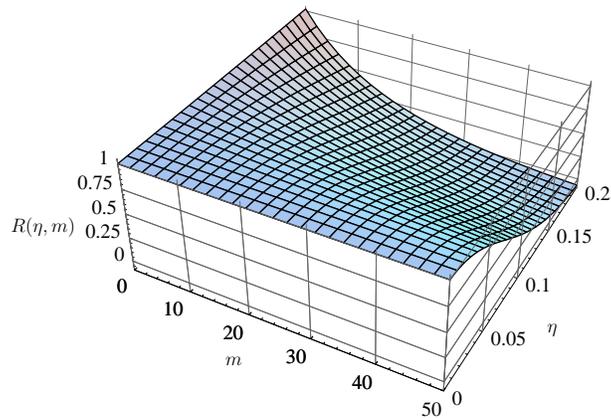}
\caption{\label{val}Ratio between the exact coupling constant and
the approximate one in the Lamb-Dicke regime. The approximation is
valid on the region $R(\eta,m)\approx 1$ where $\eta$ and $m$ are
small enough.}
\end{figure}

We would like to point out that the linear dependence of the ion-field coupling constant
on the operator $\hat{a}^{\dagger}\hat{a}$ is crucial for the generation of
the states $|\Phi_\pm\rangle_v$. Therefore, it is very important to be
aware of the limits where the parameter $\eta$ and the initial
magnitude $\alpha$ may be varied and still having the approximation
(\ref{expan}) valid. This limit is set by keeping the product
$\eta^2\overline{\hat{a}^\dagger\hat{a}}$ small enough, what allows us
to neglect higher order terms in the cosine expansion. If the
Lamb-Dicke approximation was not performed, it would be necessary to
work with the full nonlinear coupling constant $\lambda\equiv\langle
m|\cos\eta(\hat{a}^{\dagger}+\hat{a})|m\rangle=e^{-\eta^2/2}L_m^0(\eta^2)$.
For convenient values of the product $\eta^2\overline{\hat{a}^\dagger\hat{a}}$
this coupling constant reduces to $\lambda_{LD}=1-\eta^2(1+2m)/2$,
that is the coupling constant we have used so far (Lamb-Dicke regime).
In figure \ref{val} we show the ratio between the exact and the approximate
coupling constants, $R(\eta,m)\equiv\lambda/\lambda_{LD}$.
We see that there are ranges of values of $\eta$ and $m$
for which $R\approx 1$. Under such circumstances, the Lamb-Dicke
approximation is valid and the generation protocol proposed here is applicable.
\subsection{Phonon statistics and continuous observation}
\begin{figure}
\includegraphics[width=10.cm]{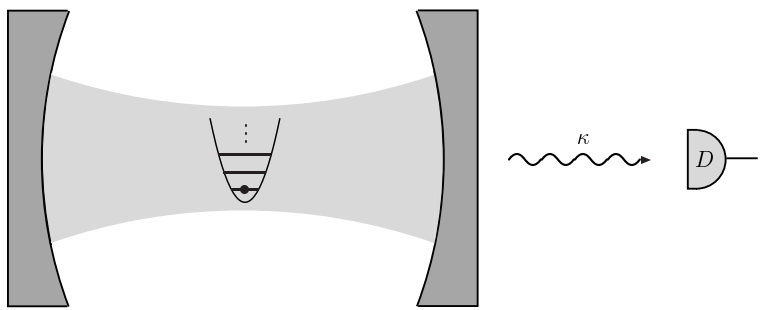}
\caption{\label{cavity}Schematic experimental setup. The system
consists of a single trapped ion placed in a lossy cavity
having a decay rate $\kappa$. The detector $D$ continuously
monitors this decay channel.}
\end{figure}

We are now interested in the more realistic setting where the cavity
is not ideal and one could detect photons leaking through its
mirrors. The set up we have in mind is depicted in Fig.\ref{cavity}.
We still have a two-level trapped ion interacting resonantly with
the cavity field but now we consider the cavity to be lossy,
decaying at a rate $\kappa$. We assume that a detector $D$ is placed outside the cavity
in a way that it may monitor the cavity decay.

\begin{figure}
\includegraphics[width=8.cm]{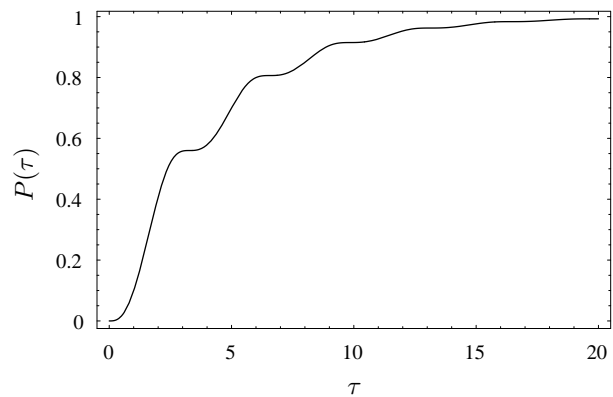}
\caption{\label{prob}Probability of detection of one photon outside the
cavity. The system parameters are $\Gamma=1$, $\eta=0.05$, and
$\alpha=2$. This probability tends to one for higher values of
$\tau$.}
\end{figure}

The description of damping in quantum optical systems is usually
described using master equations and its solution gives the time
evolution of the system when the decay is not observed. However,
the time evolution under continuous observation of photon
counts may be adequately described by a pure state that evolves according to a
non-Hermitian Hamiltonian. This approach is known as quantum jumps
or quantum trajectories \cite{qjreview} formalism. The idea of continuous
observation of decaying channels in systems consisting of atoms or
ions and cavities has proved itself to be useful to perform
legitimate information processing tasks as teleportation
\cite{martin_tele} and maximally entangled state generation \cite{martin1,martin2} or quantum gates \cite{qc_cont}, for instance. We
saw that the time evolution of the system under Hamiltonian
(\ref{hamilint}), and the realization of a measurement on the
electronic state may be used to generate the states (\ref{ent2}).
Now, instead of measuring the atomic state, we will show that a
measurement of the photon outside the cavity collapses the state of
the system in a state that keeps much of the characteristics of the
state (\ref{ent2}), namely oscillatory behavior of the distribution
$P_m$ as well as its narrowing and broadening \cite{phase}. For the sake of
simplicity, we assume that the detector $D$ is perfect.
Otherwise we would just have to account for a finite probability
that the detector fails in detecting an event of leaking of a photon,
what would lead us to a description in terms of density matrices rather than
state vectors. The time evolution of the system conditioned to a no
photon decay is given by
\begin{equation}
i\hbar\frac{d|\psi\rangle}{dt}=\hat{H}_{\rm{eff}}|\psi\rangle,\label{eq}
\end{equation}
where
\begin{eqnarray}
\hat{H}_{\rm{eff}}=-i\hbar\frac{\,\kappa\,\hat{b}^{\dagger}\hat{b}}{2}+\hbar
g \left[1-\frac{\eta^2(1+2\hat{a}^{\dagger}\hat{a})}{2}\right]
(\hat{\sigma}_- \hat{b}^{\dagger} + \hat{\sigma}_+
\hat{b}).\label{heff}\nonumber\\
\end{eqnarray}
It is worth noticing that once the Hamiltonian (\ref{heff}) is not
Hermitian, the norm of $|\psi(t)\rangle$ is not constant in time.
So, it must be normalized in order to allow one to correctly
evaluate any property of the system. It is clear that if the
initial state of the system is the same as before, namely,
$|\psi(0)\rangle=|\alpha\rangle_v|0\rangle_c|e\rangle$, the solution
of equation (\ref{eq}) may be written as
\begin{equation}
|\psi(t)\rangle=\sum_m^\infty
a_m(t)|m,0,e\rangle+b_m(t)|m,1,g\rangle\label{ev}
\end{equation}
Substituting (\ref{ev}) and (\ref{heff}) into (\ref{eq}) one obtains
two coupled differential equations that may be easily solved and the
result is given by
\begin{eqnarray}
a_m(\tau)&=& c_m(0)\,e^{-\Gamma \tau/4}\,\left(C(\tau)+\frac{\Gamma
}{\sqrt{\Gamma^2-16\lambda_{LD}^2}}\,S(\tau)\right)\nonumber\\
b_m(\tau)&=& -4\,ic_m(0)\,e^{-\Gamma
\tau/4}\,\frac{\lambda_{LD}}{\sqrt{\Gamma^2-16\lambda_{LD}^2}}\,S(\tau),
\label{coef}
\end{eqnarray}
where $c_m(0)$ are the coefficients of the expansion of the initial
coherent state in the Fock basis, $\Gamma=\kappa/g$, $\tau=gt$, and
\begin{eqnarray}
C(\tau)&=&\cosh(\sqrt{\Gamma^2-16\lambda_{LD}^2}\,\tau/4)\\
S(\tau)&=&\sinh(\sqrt{\Gamma^2-16\lambda_{LD}^2}\,\tau/4)
\end{eqnarray}
Now, we suppose that one photon is detected outside the cavity.
This event would correspond to the destruction of one photon leading
the system to state $\hat{b}|\psi(\tau)\rangle$. Again, we
remember that since the time evolution is not unitary the state must
be normalized after this jump. In our case, the resulting state
would be
$|\psi(\tau)\rangle_{d}=|\Phi(\tau)\rangle_v|0\rangle_c|e\rangle$, i.e.,
a disentangled state having a normalized motional part given
by
\begin{equation}
|\Phi(\tau)\rangle_v=\sum_{m=0}^\infty
\frac{b_m(\tau)}{\sqrt{\sum_{p=0}^\infty |b_p(\tau)|^2}}
|m\rangle_v.\label{state}
\end{equation}
Before investigating the statistical properties of that state, it is
important to calculate the probability for a photon emission because
it is related to our probability of success in generating
$|\Phi(\tau)\rangle_v$. The probability that at least one jump occurs between the initial
instant $0$ and the subsequent instant $\tau$ is given
by $P(\tau)=1-\langle \psi(\tau)|\psi(\tau)\rangle$, where
$|\psi(\tau)\rangle$ is the state in equation (\ref{ev}) with the coefficients
(\ref{coef}). In figure \ref{prob} we have a plot showing the behavior of
$P(\tau)$ using parameters that are close to the ones in a current
experimental situation, i.e. the weak coupling regime.

\begin{figure}
\includegraphics[width=8.cm]{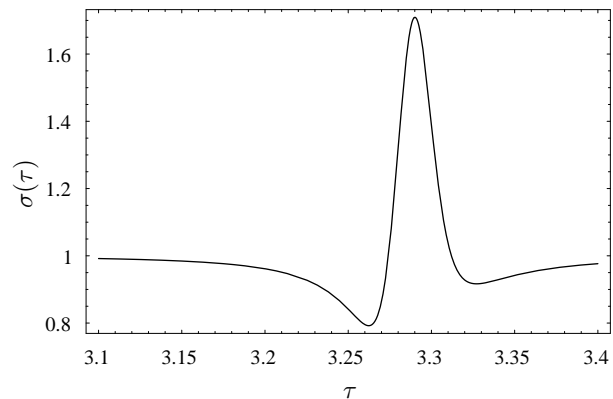}
\caption{\label{sigma}Time evolution of the normalized variance. The
system parameters are $\Gamma=1$, $\eta=0.05$, and $\alpha=2$.}
\end{figure}

\begin{figure}
\includegraphics[width=8.cm]{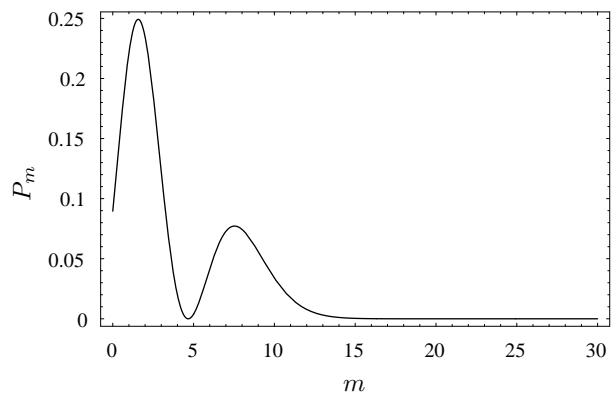}
\caption{\label{osc}Phonon distribution at $\tau=3.29$
with system parameters $\Gamma=1$, $\eta=0.05$, and $\alpha=2$.}
\end{figure}

Let us now start the analysis of the statistical properties of the
vibrational state $|\Phi(\tau)\rangle_v$. The ion started in a
coherent state which has a Poissonian distribution. We can describe
its narrowing (or widening) via the normalized variance (also know
as the {\it Fano factor}) defined as
$\sigma^2=(\overline{m^2}/\bar{m})-\bar{m}$ where $\bar{m}$ and
$\overline{m^2}$ are the first and second moments of the
distribution $P_m=|\langle m|\Phi(\tau)\pm\rangle_v|^2$,
respectively. Values of $\sigma<1$ indicate sub-Poissonian, $\sigma>
1$ super-Poissonian, and $\sigma=1$ Poissonian statistics. The time
evolution of $\sigma(\tau)$ is shown in figure \ref{sigma}. The
original Poissonian distribution naturally evolves to either sub or
super-Poissonian values. These changes in the width of the
distribution could be observed even in a bad cavity that has a decay
rate $\kappa$ comparable to the coupling constant $g$, as we can see
in figure \ref{sigma}. We would like also to show that strong
signatures of nonclassical behavior, such as the oscillations in the
phonon distribution $P_m$ at times when the statistics is
Poissonian, still persist in the weak coupling regime. This may be
seen in figure \ref{osc} where we show the distribution at a time
$\tau=3.29$ and with $\eta=0.05$.
Based on those considerations we conclude that general properties of
coherent states superpositions, which arise in the lossless case,
would still persist in our more realistic setup. This means that our proposal
could be useful for the experimental investigation of certain nonclassical features.
\section{conclusions}
We have investigated several aspects of the dynamics of a trapped
ion inside a cavity. Firstly we have considered a situation in which the
unitary time evolution leads to a global entangled state involving superposed
motional coherent states, Fock photon states and the two internal
electronic states. After the measurement of the internal
state of the ion in a specific interaction time, the generation
of quantum superposition of coherent states of motion of the ion is accomplished.
Two different states may be generated (either $|\Phi_+\rangle$ or $|\Phi_-\rangle$)
depending on the result of the measurement of the internal ionic state.
The main requirement for such generation is the strong coupling regime where
the system may perform Rabi oscillations in the lifetime of the cavity photon.
In the second part of our paper we consider the influence of cavity decay in the
ionic dynamics. In fact that represents itself a generation method, since a
nonclassical state results from the dissipative evolution even with a photon decay
rate of the same order as the ion-cavity coupling (weak coupling regime). The cavity
is continuously monitored by a detector, what
causes the state of the system to be pure at any time. The
measurement of the internal electronic state in the former
suggestion is replaced now by the counting of a photon leaking out of the
cavity. This collapses the entangled global state of the system onto
a product state. Even though the cavity is not ideal, the ionic motional state
still retains (after the photon decay) important nonclassical features that
characterize quantum superposition of coherent states, such as, for instance,
changes in the variance of the phonon distribuion (sub or super-Poissonian statistics)
as well as its oscillatory behavior.
\begin{acknowledgments}
We would like to thank Martin Plenio for reading the manuscript and
giving valuable suggestions. This work is partially supported by
CNPq (Conselho Nacional para o Desenvolvimento Cient\'\i fico e
Tecnol\'ogico), and FAPESP (Funda\c c\~ao de Amparo \`a Pesquisa do
Estado de S\~ao Paulo) grant number 02/02715-2, Brazil.
\end{acknowledgments}

\bibliography{cat}

\end{document}